\def\ANON{0} 
\def\PAGENUMS{1} 

\documentclass[conference]{IEEEtran}
\IEEEoverridecommandlockouts
\usepackage{fancyhdr} 
\usepackage{lastpage} 
\usepackage{cite}
\usepackage{amsmath,amssymb,amsfonts}
\usepackage{algorithmic}
\usepackage{booktabs}
\usepackage{multirow}
\usepackage{graphicx}
\usepackage{subcaption}
\usepackage{textcomp}
\usepackage{xcolor}
\usepackage{balance}
\usepackage[hidelinks]{hyperref}
\usepackage{xurl}
\usepackage{tabularx}
\usepackage{siunitx}
\newcommand{\sym}[1]{\textsuperscript{#1}}
\usepackage{threeparttable}
\def\BibTeX{{\rm B\kern-.05em{\sc i\kern-.025em b}\kern-.08em
    T\kern-.1667em\lower.7ex\hbox{E}\kern-.125emX}}
\begin{document}

\title{Systemic Risk in DeFi: A Network-Based Fragility Analysis of TVL Dynamics}

\ifnum\ANON=1
    \author{
        \IEEEauthorblockN{Anonymous Authors}
        \IEEEauthorblockA{\textit{} \\
        \textit{}\\
        \\
        }
    }
\else

    \author{
        \IEEEauthorblockN{
            Shiyu Zhang, 
            Zining Wang, 
            Jin Zheng and
            John Cartlidge
        }
        \IEEEauthorblockA{
            \textit{School of Engineering Mathematics and Technology}\\
            \textit{University of Bristol, Bristol, UK}\\
            \texttt{\{shiyu.zhang, zining.wang, jin.zheng, john.cartlidge\}@bristol.ac.uk}}
    }

\fi
\maketitle

\ifnum\PAGENUMS=1
    \thispagestyle{fancy}
    \pagestyle{fancy}
    \fancyfoot[C]{\fontsize{9}{10} \selectfont Page \thepage ~of {\hypersetup{hidelinks}\pageref{LastPage}}}
\fi

\begin{abstract}
Systemic risk refers to the overall vulnerability arising from the high degree of interconnectedness and interdependence within the financial system. In the rapidly developing decentralized finance (DeFi) ecosystem, numerous studies have analyzed systemic risk through specific channels such as liquidity pressures, leverage mechanisms, smart contract risks, and historical risk events. However, these studies are mostly event-driven or focused on isolated risk channels, paying limited attention to the structural dimension of systemic risk. Overall, this study provides a unified quantitative framework for ecosystem-level analysis and continuous monitoring of systemic risk in DeFi. From a network-based perspective, this paper proposes the DeFi Correlation Fragility Indicator (CFI), constructed from time-varying correlation networks at the protocol category level. The CFI captures ecosystem-wide structural fragility associated with correlation concentration and increasing synchronicity. Furthermore, we define a Risk Contribution Score (RCS) to quantify the marginal contribution of different protocol types to overall systemic risk. By combining the CFI and RCS, the framework enables both the tracking of time-varying systemic risk and identification of structurally important functional modules in risk accumulation and amplification.
\end{abstract}


\section{Introduction}
\noindent
Decentralized finance (DeFi) has developed rapidly in recent years, evolving from a relatively isolated set of protocols into an on-chain financial ecosystem with economic activity reaching hundreds of billions of U.S. dollars at its peak, and encompassing a wide range of financial functions~\cite{leon2025data}. As the number of DeFi protocols grows and composability becomes a defining design principle, interdependencies across different protocols types have intensified~\cite{schar2021decentralized,werner2022sok}. In this environment, shocks originating from a single protocol or market segment may propagate through shared collateral, common units of account, and tightly coupled smart contracts, amplifying localized disturbances into system-wide stress~\cite{board2023financial}. Compared with traditional financial systems, the programmable and composable nature of DeFi renders these propagation channels more explicit yet potentially more fragile, making systemic risk particularly salient in this environment.

A large literature in traditional finance establishes that systemic risk is fundamentally a network phenomenon, arising from patterns of interdependence rather than the size of individual institutions~\cite{haldane2011systemic,bardoscia2017pathways}. Correlation-based network representations have been widely used to uncover latent market structures and collective dynamics~\cite{mantegna1999hierarchical}, and to develop quantitative measures of interconnectedness and systemic risk~\cite{billio2012econometric,diebold2014network}. Subsequent work emphasizes that network topology, density, and synchronization shape systemic vulnerability, showing that increased connectivity can amplify fragility during stress periods~\cite{acemoglu2015systemic,battiston2012debtrank}. These insights motivate the use of structural and time-varying network indicators for systemic risk monitoring.

As crypto-asset markets and DeFi have matured, systemic risk research has increasingly extended to on-chain environments~\cite{aquilina2025cryptocurrencies}. Existing studies highlight several features that distinguish DeFi from traditional systems and discuss their impact on exacerbating systemic vulnerability, including protocol composability~\cite{kitzler2023disentangling}, shared collateral and collateral reuse~\cite{qin2021attacking}, and automated liquidation mechanisms that create endogenous feedback between prices and liquidations~\cite{auer2018regulating}. While this literature provides important insights into specific risk mechanisms, it typically focuses on individual protocols, assets, or crisis episodes, offering limited tools for unified and time-consistent ecosystem-level risk measurement.

Related work applies network-based methods to study dependence structures and shock propagation in crypto and DeFi markets. Correlation and dependence networks constructed from prices or on-chain activity reveal that interconnections strengthen during stress periods, signaling elevated systemic vulnerability~\cite{fakhfekh2024dependence,feng2025research,yan2025network}. Other studies construct explicit exposure networks or simulate contagion through liquidation cascades~\cite{wu2025dexposure,tovanich2023contagion}. Although these methods are valuable in revealing specific risk transmission mechanisms, their analysis often relies on scenario assumptions or model settings, focusing more on local shocks or short-term propagation processes, making them difficult to use for continuous and comparable structural monitoring of systemic risk in DeFi.

Overall, the literature provides rich insights into the mechanisms and transmission channels through which systemic risk emerges in DeFi, yet remains fragmented in how such risks are summarized and monitored at the ecosystem level. Most existing approaches focus on localized mechanisms, specific assets, or crisis episodes, and therefore offer limited guidance on how structural vulnerabilities accumulate and persist over time. In particular, the literature lacks a measurement perspective that treats systemic risk as a dynamic structural state—one that can be tracked continuously, compared across market phases, and decomposed across functional modules.

To address these limitations, we adopt a network-based perspective and conceptualize systemic risk in DeFi as a structural state variable that evolves over time and can be continuously monitored at the ecosystem level. We construct time-varying correlation networks at the protocol category level to capture evolving interdependencies and synchronization across functional modules, moving beyond scenario-driven or event-specific analyzes. Based on these networks, we propose the Correlation Fragility Indicator (CFI), which provides a time-consistent and cross-period comparable measure of ecosystem-wide structural vulnerability implied by increasing correlation concentration and synchronization. To link system-level risk measurement with its structural origins, we further introduce the Risk Contribution Score (RCS), which decomposes changes in systemic risk into marginal contributions from functional modules, enabling structural attribution of risk accumulation and amplification across the DeFi ecosystem.

The main contributions of this paper are:
\begin{itemize}
    \item We propose a unified analytical framework that treats systemic risk in DeFi as a structural and time-evolving state variable for continuous ecosystem-level monitoring.
    \item We develop CFI, a time-consistent and cross-period comparable measure that captures ecosystem-wide structural vulnerability arising from evolving dependency and synchronization patterns.
    \item We propose the RCS, which structurally decomposes system-level risk into marginal contributions of different functional modules, providing a module-level attribution of systemic risk.
    \item Using on-chain data, we conduct an empirical analysis that documents the time-varying nature of systemic risk in DeFi and identifies protocol categories that play a critical role in risk accumulation and amplification.
\end{itemize}

\noindent\textbf{Data and Code Availability.}
Data and code used in this study are available in an anonymous GitHub repository for  review.\footnote{https://github.com/defiresearchanonymous/defi-systemic-risk}

\section{Data and Preprocessing}
\subsection{Data Source and Sample Construction}
\noindent
We construct a comprehensive dataset of Total Value Locked (TVL) for the DeFi ecosystem using the public API provided by DeFiLlama.\footnote{https://defillama.com} TVL measures the aggregate value of assets deposited in DeFi protocols and is widely used as a core indicator of capital allocation and economic activity. The raw data consist of daily U.S. dollar–denominated TVL observations at the protocol level, covering more than 5,000 DeFi protocols. 

To analyze the structural evolution of the ecosystem, we map each protocol to one of the standardized protocol types provided by DeFiLlama (e.g., \textit{Lending}, \textit{DEX}, \textit{Liquid Staking}). Protocols categorized as \textit{CEX} and \textit{Chain} are excluded because they do not represent DeFi-native activities. After filtering and aggregation, the final dataset spans 2021-01-01 to 2025-10-31 and comprises 70 protocol types.\footnote{https://defillama.com/categories} For each category, daily TVL is obtained by summing the TVL of all constituent protocols, yielding a balanced panel in which each node corresponds to a DeFi protocol type.

\subsection{Data Cleaning and Anomaly Detection}
\noindent
Raw DeFi TVL data contain substantial irregularities due to oracle disruptions, API glitches, contract upgrades, or temporary reporting errors. To mitigate such noise while preserving true market dynamics, we apply a two-step anomaly detection procedure based on relative and absolute daily TVL changes and deviations from the median absolute deviation (MAD). Specifically, an observation at time $t$ is flagged as anomalous if it satisfies any of the following conditions:

\begin{equation}
\begin{aligned}
|\Delta \mathrm{TVL}_t| &> \tau_{\text{abs}}, \\
\left|\frac{\mathrm{TVL}_t}{\mathrm{TVL}_{t-1}} - 1\right| &> \tau_{\text{rel}}, \\
\frac{|\Delta \mathrm{TVL}_t - \text{median}(\Delta \mathrm{TVL})|}
{\text{MAD}(\Delta \mathrm{TVL})} &> \tau_{\text{MAD}}.
\end{aligned}
\end{equation}

We set $\tau_{\text{abs}} = 5$ million USD, $\tau_{\text{rel}} = 200\%$, and $\tau_{\text{MAD}} = 12$. In practice, 3.31\% of protocol-day observations are flagged as technical anomalies, typically corresponding to abrupt spikes followed by immediate reversals, and are repaired via local interpolation, whereas only 0.34\% are classified as large, persistent changes and retained as economically meaningful liquidity shifts.

\subsection{Return Construction and Winsorization}
\noindent
Following standard practice in financial time series analysis, we construct daily log returns from category-level TVL:
\begin{equation}
    r_{i,t} = \log\left(\mathrm{TVL}_{i,t} + \varepsilon \right) - \log\left(\mathrm{TVL}_{i,t-1} + \varepsilon \right),
\end{equation}
where \(\varepsilon = 10^{-11}\) ensures numerical stability. Remaining gaps are filled using forward and backward interpolation, and returns are symmetrically winsorized at the 0.5\% level to mitigate the undue influence of extreme observations. Categories with zero return variance are excluded, yielding a balanced panel of 70 categories. This processed return matrix serves as the input for constructing dynamic correlation networks in the subsequent analysis.


\section{Network Construction}
\subsection{Correlation Estimation}
\noindent
To capture cross-category dependence, we estimate correlations based on daily TVL log returns, which proxy the synchronization of capital flows in the DeFi ecosystem. However, direct sample correlations are often unstable in this setting due to volatile on-chain activity, heterogeneous liquidity across protocols, and the limited number of observations available within each rolling window. We therefore employ the Ledoit–Wolf shrinkage estimator~\cite{ledoit2004well}, which is widely used in high-dimensional financial return data~\cite{demiguel2009optimal,ledoit2017nonlinear}.

Let $r_{i,t}$ denote the log return of category $i$ on date $t$, and let $S_t$ be the sample covariance matrix computed over the rolling window ending at~$t$. The shrinkage estimator is given by
\begin{equation}
    \hat{\Sigma}_t = (1-\lambda_t) S_t + \lambda_t T_t,
\end{equation}
where $T_t = \mu_t I$ is a scaled identity matrix, with $I$ denoting the identity matrix and $\mu_t$ equal to the average variance (i.e., the mean diagonal element of $S_t$). The shrinkage intensity $\lambda_t \in [0,1]$ is selected to minimize expected mean-squared error. The correlation matrix is then obtained by standard normalization
\begin{equation}
    C_{ij,t} = 
    \frac{\hat{\Sigma}_{ij,t}}
    {\sqrt{\hat{\Sigma}_{ii,t}\,\hat{\Sigma}_{jj,t}}}.
\end{equation}

These shrinkage-based correlation matrices provide stable and economically interpretable dependence estimates and form the basis of our network analysis. Robustness to alternative network specifications is examined in Section~\ref{sec:robustness_alt_spec}.

\subsection{Rolling Window Framework}
\noindent
Empirical evidence suggests that dependence among DeFi protocols is highly time-varying, driven by fluctuations in user activity, liquidity migration, and market-wide shocks. A single full-sample correlation matrix would obscure these temporal dynamics and fail to capture periods of elevated synchronization or structural breaks~\cite{engle2002dynamic,forbes2002no}. 

To capture these dynamics, we adopt a rolling-window estimation scheme, computing shrinkage correlation matrices over overlapping windows of length $W=120$ days with a step size of $\Delta=7$ days. This choice, guided by sensitivity analysis, balances smoothness and responsiveness, allowing the network to track gradual expansions and contractions in connectivity while remaining sensitive to sustained co-movement during stress periods.



\subsection{Network Definition}
\noindent
Given the shrinkage correlation matrix $C_{t}$ for each rolling window, we construct a weighted, undirected network to capture co-movement among protocol types. For each window ending at time $t$, let $C_{ij,t}$ denote the correlation between types $i$ and $j$. We construct a weighted adjacency matrix $A_t = (w_{ij,t})$ applying a monotone transformation of the correlations:
\begin{equation}
    w_{ij,t} = | C_{ij,t} |,
\end{equation}
for all $i \neq j$, and $w_{ii,t} = 0$. 

Using absolute correlations ensures that both positive and negative co-movements are treated as potential channels of risk transmission, consistent with systemic risk applications where the magnitude of dependence is of primary interest~\cite{onnela2003dynamics}. The resulting network is fully connected and weighted, allowing us to retain the complete dependence structure among protocol types types and avoiding arbitrary correlation thresholds that may distort global network measures.

\begin{figure*}[t]
    \centering
    \includegraphics[width=0.9\textwidth]{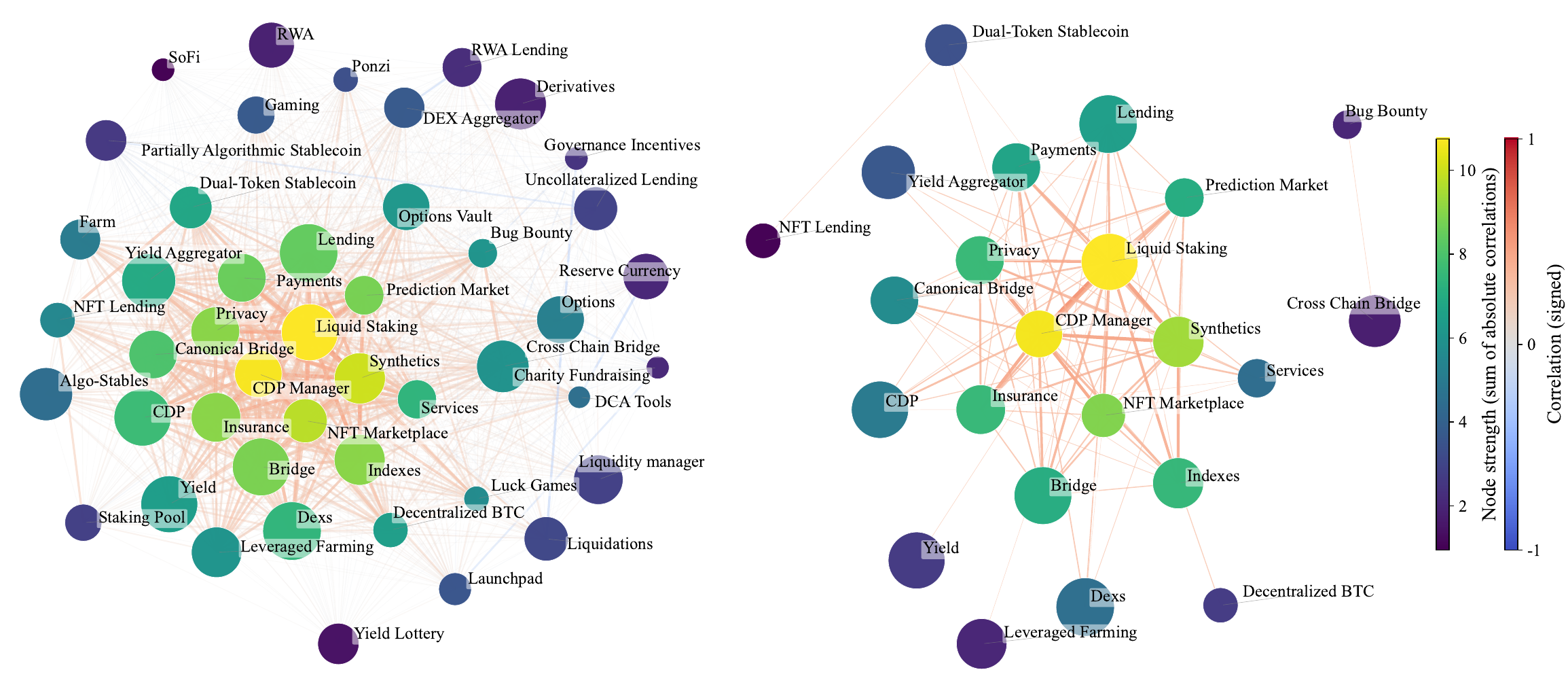}
    \caption{DeFi correlation networks (snapshot at rolling-window end date: 11 May 2022). 
    The left panel shows the fully weighted correlation network constructed from category-level TVL log returns over the rolling window, where each edge weight equals the absolute correlation $|C_{ij}|$. 
    The right panel shows the thresholded network retaining only edges with $|C_{ij}| > 0.3$ to highlight the core of strong dependencies. 
    Node size is proportional to category-level TVL, node color denotes node strength (sum of absolute correlations), edge color indicates signed correlation, and edge width reflects correlation magnitude.}
    \label{fig:network_snapshot}
\end{figure*}

For transparency and robustness, we also consider thresholded networks that retain only strong dependencies. These networks are used for visualization and robustness analysis, while all baseline systemic risk measures rely on the fully connected weighted network. Fig.~\ref{fig:network_snapshot} shows a representative snapshot, revealing a clear core–periphery structure: strong dependencies cluster around major DeFi functions such as lending and liquid staking, whereas peripheral types drop out under thresholding. Notably, the persistent centrality of CDP management and synthetic asset types highlights less obvious but systemically important channels of interconnectedness.

\section{System-Level: Correlation Fragility Indicator (CFI)}
\label{sec:cfi}
\noindent
Here, we develop a network-based systemic risk indicator, termed CFI, to capture the structural fragility of the DeFi ecosystem. Unlike conventional risk measures centered on returns~\cite{leon2025data}, prices~\cite{schar2021decentralized}, or event-driven shocks~\cite{board2023financial}, CFI quantifies the evolving dependence structure embedded in time-varying correlation networks. By construction, it reflects how tightly coupled the ecosystem becomes, capturing persistent co-movements in liquidity across protocol categories as measured by rolling-window correlation networks spanning several weeks to months, rather than short-lived transitory shocks.

\subsection{Network Metrics and Design Principles}
\label{subsec:cfi_metrics}
\noindent
To quantify structural fragility in correlation networks, we select four complementary metrics grounded in the spectral and network analysis of financial correlation matrices. Together, they capture key dimensions of dependence concentration and diversification, including system-wide synchronization, dominance of common factors, prevalence of extreme bilateral linkages, and dispersion of the dependence spectrum~\cite{tumminello2007correlation}. This yields a parsimonious and economically interpretable representation of correlation-network fragility.

First, \emph{average strength} summarizes the overall intensity of pairwise dependence across protocol types and captures system-wide synchronization~\cite{barrat2004architecture}:
\begin{equation}
\bar{s}_t=\frac{1}{N}\sum_{i=1}^N \sum_{j\neq i} w_{ij,t}.
\end{equation}

Second, the \emph{maximum eigenvalue} of the weighted adjacency correlation matrix captures the dominance of a common latent factor in the dependence structure:
\begin{equation}
\lambda_{\max,t}=\max\{\lambda_k(A_t)\}.
\end{equation}
In correlation-based financial networks, the largest eigenvalue is commonly interpreted as a market-wide or systemic mode reflecting collective behavior~\cite{laloux1999noise,plerou2002random}.

Third, \emph{strong-edge density} measures the prevalence of unusually strong bilateral dependencies by computing the fraction of correlations exceeding a given threshold:
\begin{equation}
d^{\text{strong}}_t=\frac{1}{N(N-1)}\sum_{i\neq j}\mathbf{1}\{|C_{ij,t}|>\rho\},
\quad \rho=0.3,
\end{equation}
where $\rho=0.3$ reflects a conservative cutoff for economically meaningful co-movements; robustness to alternative thresholds is assessed in Section~\ref{sec:robust_threshold}. This metric captures whether strong co-movements form a tightly connected core, consistent with filtered correlation-network approaches~\cite{tumminello2005tool}.

Finally, \emph{eigenvalue entropy} summarizes the dispersion of the eigenvalue spectrum and provides a compact measure of diversification in network dependence:
\begin{equation}
H_t=-\frac{1}{\log N}\sum_{k=1}^N p_{k,t}\log p_{k,t},
\quad
p_{k,t}=\frac{\lambda_{k,t}}{\sum_{j=1}^N \lambda_{j,t}}.
\end{equation}
Lower entropy indicates a more concentrated spectrum and thus a more fragile dependence structure~\cite{kwapien2012physical}.

Taken together, higher average strength, a larger maximum eigenvalue, greater strong-edge density, and lower eigenvalue entropy consistently indicate a more tightly coupled and less diversifiable correlation network.

\subsection{Construction of the CFI}
\label{subsec:cfi_construction}
\noindent
To summarize multiple dimensions of correlation-network fragility into a single and interpretable measure, we construct the DeFi CFI using Principal Component Analysis (PCA). PCA-based aggregation of co-moving stress indicators is standard in systemic risk measurement, most notably in the construction of composite financial stress indices \cite{hollo2012ciss}. By design, the CFI captures the dominant common variation across network metrics that reflect synchronization, concentration, and diversification in the dependence structure.

 Let $x_{k,t}$ denote metric $k\in\{1,\ldots,4\}$ at window $t$. Each metric is computed over a rolling window and assigned to the corresponding window end date, so that $x_{k,t}$ captures the average structural properties of the correlation network over that window rather than instantaneous fluctuations. Each metric is then standardized to ensure comparability across dimensions,
\begin{equation}
\tilde{x}_{k,t}=\frac{x_{k,t}-\mu_k}{\sigma_k},
\end{equation}
where $\mu_k$ and $\sigma_k$ are the sample mean and standard deviation of metric $k$. Let $\tilde{\mathbf{x}}_t=(\tilde{x}_{1,t},\ldots,\tilde{x}_{4,t})^\top$ collect the standardized metrics. We then apply PCA to the covariance matrix of $\tilde{\mathbf{x}}_t$ and define the CFI as the first principal component:
\begin{equation}
\mathrm{CFI}_t=\mathbf{w}_1^\top \tilde{\mathbf{x}}_t,
\end{equation}
where $\mathbf{w}_1$ is the eigenvector associated with the largest eigenvalue. By construction, the CFI represents the dominant common component underlying the four fragility dimensions. The sign of the component is oriented such that higher values correspond to stronger network-wide synchronization, i.e., a positive association with average strength. To facilitate interpretation and comparability across time and empirical specifications, the resulting CFI series is subsequently standardized to have zero mean and unit variance.

Table~\ref{tab:cfi_pca} reports the PCA diagnostics underlying the construction of the CFI. The first principal component explains an overwhelming share of the joint variation in the four standardized metrics, and its loadings exhibit a coherent and economically intuitive pattern: average strength, maximum eigenvalue, and strong-edge density load positively, while eigenvalue entropy loads negatively. This empirical structure confirms that the CFI effectively aggregates the core dimensions of correlation-network fragility into a single state variable.

\begin{table}[t]
\centering
\caption{PCA Diagnostics for the CFI Construction}
\label{tab:cfi_pca}
\begin{tabular}{
    l
    S[table-format=+1.4]
    S[table-format=1.4]
}
\toprule
 & {PC1 loading} & {{PC1 variance share}} \\
\midrule
Average strength & 0.5054 & \multicolumn{1}{c}{\multirow{4}{*}{0.9673}} \\
Maximum eigenvalue & 0.5052 &  \\
Strong-edge density ($|C_{ij}|>0.3$) & 0.4983 &  \\
Eigenvalue entropy & -0.4909 &  \\
\bottomrule
\end{tabular}
\end{table}

\subsection{Dynamics and Risk-Monitoring Relevance}
\label{subsec:cfi_descriptive}
\begin{figure}[t]
\centering
\includegraphics[width=0.95\linewidth]{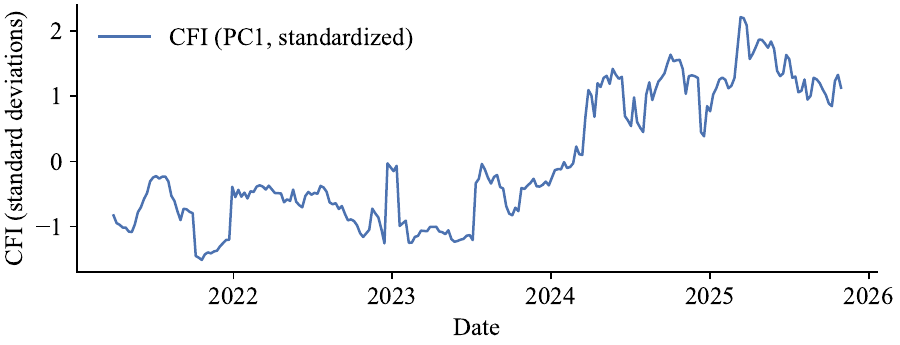}
\caption{Time series of the standardized DeFi CFI based on rolling correlation networks of category-level TVL log returns. The series is oriented so that higher values indicate stronger network-wide synchronization and higher structural fragility.}
\label{fig:cfi_timeseries_full70}
\end{figure}

\noindent
Fig.~\ref{fig:cfi_timeseries_full70} plots the standardized CFI over time. Each observation corresponds to the end date of a rolling estimation window and summarizes the dependence structure of the DeFi ecosystem within that window. The CFI exhibits pronounced medium-run variation and evolves smoothly across time, reflecting the gradual accumulation and release of structural dependence in the correlation network. This smooth evolution is consistent with the interpretation of the CFI as a structural state variable capturing persistent shifts in ecosystem-wide synchronization, rather than high-frequency market fluctuations.

To assess whether the CFI tracks contemporaneous systemic conditions, we relate it to a set of realized risk proxies via,
\begin{equation}
\text{Risk}_t=\alpha+\beta\,\mathrm{CFI}_t+\gamma^\top \mathbf{Z}_t+\varepsilon_t,
\end{equation}
where $\text{Risk}_t$ denotes a contemporaneous risk proxy and $\mathbf{Z}_t$ includes standard market controls. While the CFI is constructed using a 120-day rolling window, the dependent variables are measured over shorter horizons to reflect realized risk conditions at the same time. Inference is based on heteroskedasticity and autocorrelation consistent (HAC) standard errors.

Table~\ref{tab:risk_monitoring} shows that the CFI is not significantly associated with contemporaneous ETH volatility once controls are included, indicating that it does not merely proxy short-term price-based risk. In contrast, the CFI is positively associated with rolling aggregate TVL volatility, suggesting that periods of elevated correlation fragility coincide with heightened system-wide liquidity instability. Overall, the results highlight the distinct informational content of the CFI as a network-based structural indicator, capturing dimensions of systemic vulnerability not fully reflected in asset price dynamics.

\begin{table}[t]
\centering
\caption{Risk Monitoring Regressions: Contemporaneous Systemic Conditions}
\label{tab:risk_monitoring}
\footnotesize
\begin{threeparttable}
\setlength{\tabcolsep}{5pt}
\renewcommand{\arraystretch}{1.1}
\begin{tabular}{l
S[table-format=-2.3, table-space-text-post=\sym{***}]
S[table-format=1.4,  table-space-text-post=\sym{***}]}
\toprule
& \multicolumn{1}{c}{(1)} & \multicolumn{1}{c}{(2)} \\
& \multicolumn{1}{c}{ETH Volatility$_t$} & \multicolumn{1}{c}{TVL Volatility$_t$} \\
\midrule
CFI$_t$
& -0.026 & 0.0020\sym{*} \\
& (0.060) & (0.0011) \\
ETH Volatility$_t$ (control)
& 1.574\sym{**} & 0.0069 \\
& (0.707) & (0.0095) \\
BTC Return$_t$
& -2.199\sym{*} & {} \\
& (1.245) & {} \\
Gas Fee$_t$
& -18.348 & 0.338 \\
& (12.219) & (0.219) \\
Constant
& 0.307\sym{***} & 0.0224\sym{***} \\
& (0.076) & (0.0017) \\
\midrule
Observations & \multicolumn{1}{c}{236} & \multicolumn{1}{c}{240} \\
$R^2$ & \multicolumn{1}{c}{0.138} & \multicolumn{1}{c}{0.044} \\
HAC SE & \multicolumn{1}{c}{Yes} & \multicolumn{1}{c}{Yes} \\
\bottomrule
\end{tabular}
\vspace{0.4em}
\begin{minipage}{0.95\linewidth}
\footnotesize
\textit{Notes}: Dependent variables capture contemporaneous systemic conditions. ETH volatility is annualized 30-day realized volatility of ETH returns; TVL volatility is the 30-day rolling volatility of aggregate TVL growth. CFI is the standardized DeFi Correlation Fragility Indicator constructed from correlation networks estimated over a 120-day rolling window and evaluated at the window end date. All regressions use HAC standard errors. \sym{***}, \sym{**}, and \sym{*} denote statistical significance at the 1\%, 5\%, and 10\% levels, respectively.
\end{minipage}
\end{threeparttable}
\end{table}

To assess whether the CFI contains forward-looking information beyond contemporaneous conditions, we conduct a lightweight predictive check relating the CFI to future realized volatility of aggregate TVL growth. For horizons $h\in\{7,14,30\}$, we estimate
\begin{equation}
\label{eq:cfi_predictive_fwdvol}
V^{(h)}_t
= \alpha_h + \beta_h \,\mathrm{CFI}_t
+ \delta_h \, V^{(h)}_{t-1}
+ \gamma_h^\top \mathbf{Z}_t
+ \varepsilon_{t,h},
\end{equation}
where $V^{(h)}_t$ denotes the realized volatility of aggregate TVL log growth over the next $h$ days. 
The control vector $\mathbf{Z}_t$ includes standard market controls, and inference uses HAC standard errors with Newey--West lag length set to $2h$.

We find that the CFI is positive and statistically significant across horizons, even after controlling for volatility persistence. The estimated coefficients $\hat\beta_h$ are $0.0028$ ($h=7$), $0.0017$ ($h=14$), and $0.0010$ ($h=30$), all significant at the 1\% level with $n=239$ observations. These results indicate that elevated correlation fragility precedes increases in system-wide liquidity instability, supporting the CFI as a slow-moving structural state variable rather than a purely contemporaneous risk proxy.

\section{Node-Level: Risk Contribution Score (RCS)}
\label{sec:rcs}
\noindent
While CFI summarizes ecosystem-wide synchronization and structural fragility, it does not reveal how this fragility is distributed across protocol categories. For risk monitoring and stress testing, identifying structurally important nodes is therefore essential. This section moves from the network level to the node level and introduces RCS, a counterfactual measure of each category’s marginal contribution to the overall CFI.

\subsection{Definition of RCS}
\label{subsec:rcs_definition}
\noindent
Let $\mathcal{C}(\cdot)$ denote the fixed CFI mapping established in Section~\ref{subsec:cfi_construction}, which transforms a network snapshot into the scalar CFI using (i) the same four network metrics, (ii) the same standardization constants $(\mu_k,\sigma_k)$, and (iii) the same PCA loading vector $\mathbf{w}_1$ estimated once in the main pipeline.

For each node $i\in\{1,\ldots,N\}$, define the counterfactual network obtained by removing node $i$ (and all incident edges) from the window-$t$ network:
\begin{equation}
    A_t^{(-i)} \in \mathbb{R}^{(N-1)\times (N-1)}.
\end{equation}
We compute the counterfactual fragility state as
\begin{equation}
    \mathrm{CFI}_t^{(-i)} \equiv \mathcal{C}\!\left(A_t^{(-i)}\right),
\end{equation}
using the same fixed mapping $\mathcal{C}(\cdot)$ to ensure comparability across $i$ and across $t$.

We define the node-level RCS as the marginal contribution of node $i$ to system-wide fragility:
\begin{equation}
\label{eq:rcs_def}
    \mathrm{RCS}_{i,t}
    \equiv
    \mathrm{CFI}_t-\mathrm{CFI}_t^{(-i)}.
\end{equation}
Under our sign convention, $\mathrm{RCS}_{i,t}>0$ means that removing category $i$ reduces the CFI, implying that category $i$ increases ecosystem-wide synchronization and fragility in window $t$. Conversely, $\mathrm{RCS}_{i,t}<0$ indicates a stabilizing role.

For scale-free comparisons across system states, we also report
\begin{equation}
\label{eq:rcs_rel}
    \mathrm{RCS}^{\mathrm{rel}}_{i,t}
    =
    \frac{\mathrm{RCS}_{i,t}}{|\mathrm{CFI}_t|+\varepsilon},
\end{equation}
where $\varepsilon>0$ is a small constant to avoid division by zero.

To produce stable rankings, we aggregate RCS over time:
\begin{equation}
\label{eq:rcs_timeavg}
    \mathrm{RCS}_i = \frac{1}{|\mathcal{T}|}\sum_{t\in\mathcal{T}} \mathrm{RCS}_{i,t}.
\end{equation}
We also compute a high-fragility ranking focusing on windows in the upper tail of the CFI distribution:
\begin{equation}
\begin{aligned}
\label{eq:rcs_high}
    \mathrm{RCS}^{\mathrm{High}}_i
    =
    \frac{1}{|\mathcal{T}_{\mathrm{High}}|}\sum_{t\in\mathcal{T}_{\mathrm{High}}}
    \mathrm{RCS}_{i,t},
    \\
    \mathcal{T}_{\mathrm{High}}=\{t:\mathrm{CFI}_t \ge q_{0.75}(\mathrm{CFI})\}.
    \end{aligned}
\end{equation}

\begin{figure}[t]
    \centering
    
    \begin{subfigure}[t]{0.48\linewidth}
        \centering
        \includegraphics[width=\linewidth]{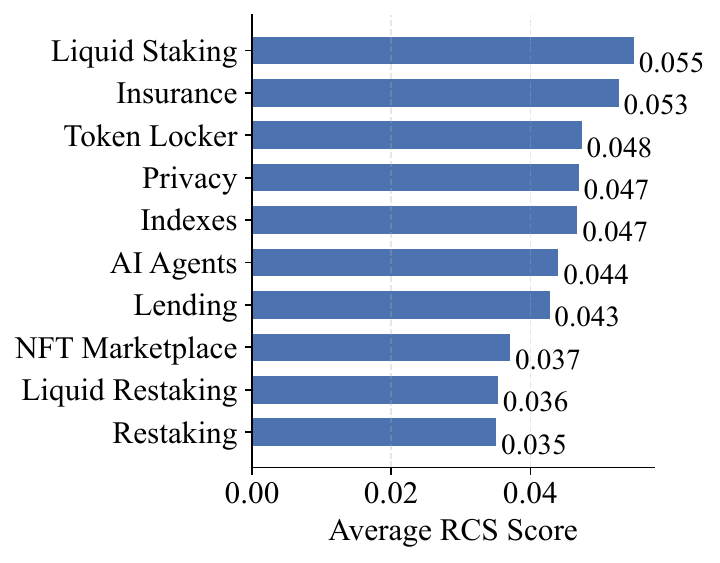}
        \caption{Top-10 protocol categories ranked by average RCS.}
        \label{fig:top10_mean_rcs}
    \end{subfigure}
    \hfill
    \begin{subfigure}[t]{0.48\linewidth}
        \centering
        \includegraphics[width=\linewidth]{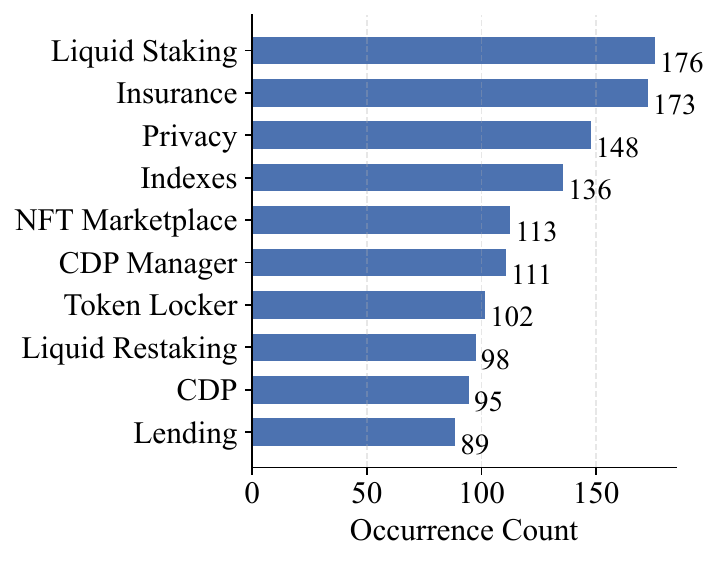}
        \caption{Appearance frequency of top-10 RCS rankings.}
        \label{fig:top10_frequency_rcs}
    \end{subfigure}
    
    \caption{Systemically important protocol categories based on RCS. Panel (a) ranks protocol types by their average marginal contribution to system-wide fragility. Panel (b) reports the frequency with which each category appears in the top-10 RCS ranking across rolling windows.}
    \label{fig:rcs_top10_combined}
\end{figure}

\subsection{Empirical Properties and Rankings}
\label{subsec:rcs_empirical}
\noindent
We next examine the identity and stability of systemically important protocol categories using the RCS. 

Fig.~\ref{fig:top10_mean_rcs} ranks protocol types by their time-averaged RCS (Eq.~\ref{eq:rcs_timeavg}), capturing their average marginal contribution to system-wide fragility. Several economically central categories consistently emerge as highly important, yet these need not be the largest by TVL (see TABLE~\ref{tab:rcs_vs_tvl_top10}), underscoring that node-level systemic importance is driven by dependence structure rather than economic size. Fig.~\ref{fig:top10_frequency_rcs} complements this static ranking by reporting how frequently each protocol type appears in the top-10 RCS ranking across rolling windows. High re-occurrence frequencies indicate that systemic importance is a persistent structural feature rather than a transient artifact of short-lived shocks. Taken together, the two panels provide a joint characterization of node-level systemic importance in terms of both magnitude and temporal stability.

Finally, we test whether structural importance simply proxies for economic scale. Table~\ref{tab:rcs_vs_tvl_top10} compares the top-10 categories by RCS with their TVL ranks and their ranks under high-fragility conditions (Eq.~\ref{eq:rcs_high}). The divergence between RCS-based ranks and TVL ranks indicates that systemic importance is not reducible to balance-sheet size. Moreover, the high-fragility ranking highlights state-dependence: stress episodes reshape the hierarchy of importance rather than uniformly amplifying existing ranks. 

\begin{table}[t]
\centering
\caption{Structural Importance versus Economic Scale}
\label{tab:rcs_vs_tvl_top10}
\footnotesize
\begin{tabular}{lccc}
\toprule
Category & Rank$_{\text{RCS}}$ & Rank$_{\text{TVL}}$ & Rank$_{\text{RCS, High}}$ \\
\midrule
Liquid Staking        & 1  & 3 & 1 \\
Insurance             & 2  & 30 & 3 \\
Token Locker          & 3  & 28 & 4 \\
Privacy               & 4  & 27 & 6 \\
Indexes               & 5  & 20 & 8 \\
AI Agents             & 6  & 35 & 14 \\
Lending               & 7  & 1 & 10 \\
NFT Marketplace       & 8  & 39 & 5 \\
Liquid Restaking      & 9  & 10 & 2 \\
Restaking             & 10 & 5 & 13 \\
\bottomrule
\end{tabular}
\end{table}

\subsection{Attack Tests: Counterfactual De-Risking via Targeted Node Removal}
\label{subsec:attack_test_rcs}
\noindent
The RCS is designed to identify protocol types that most sustain ecosystem-wide fragility. We validate this interpretation via counterfactual attack tests that quantify how much fragility can be reduced by removing a small number of categories.

For each rolling window $t$, we remove $k\in\{1,3,5,10\}$ nodes and recompute the CFI using the same fixed mapping $\mathcal{C}(\cdot)$. The de-risking effect is defined as
\begin{equation}
\label{eq:attack_drop}
\Delta \mathrm{CFI}_{t}(k) \;=\; \mathrm{CFI}_{t}^{\text{full}} \;-\; \mathrm{CFI}_{t}^{(-k)},
\end{equation}
where $\mathrm{CFI}_{t}^{(-k)}$ denotes the CFI after removing $k$ nodes.

\begin{figure}[t]
    \centering
    \includegraphics[width=0.98\linewidth]{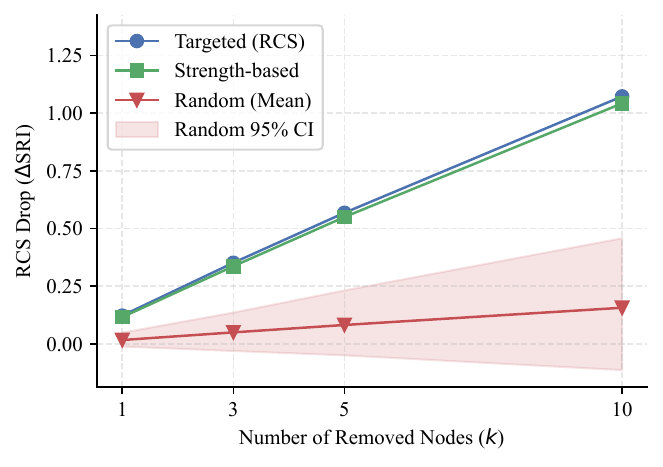}
    \caption{Attack curves: average CFI drop after removing $k$ nodes under targeted (RCS-based), strength-based, and random removal (random: 95\% interval across dates).}
    \label{fig:attack_curves_rcs}
\end{figure}

We compare three removal rules: (i) \emph{Targeted}, removing the top-$k$ categories by date-specific RCS; (ii) \emph{Strength-based}, removing the top-$k$ categories by node strength; and (iii) \emph{Random}, removing $k$ categories uniformly at random (Monte Carlo). Fig.~\ref{fig:attack_curves_rcs} shows that targeted removal produces substantially larger reductions in CFI than random removal across all $k$, and the gap widens with $k$. Table~\ref{tab:attack_summary_by_k_rcs} summarizes average effects and quantifies this excess de-risking impact through the Targeted$-$Random difference.

\begin{table}[t]
\centering
\caption{Attack Test Summary: Mean CFI Drop by Removal Size $k$}
\label{tab:attack_summary_by_k_rcs}
\footnotesize
\begin{tabular}{lcccc}
\toprule
$k$ & Targeted (RCS) & Strength & Random & Targeted $-$ Random \\
\midrule
1  & 0.122 & 0.116 & 0.017 & 0.106 \\
3  & 0.351 & 0.336 & 0.050 & 0.302 \\
5  & 0.568 & 0.550 & 0.082 & 0.486 \\
10 & 1.072 & 1.042 & 0.157 & 0.916 \\
\bottomrule
\end{tabular}

\vspace{0.35em}
\begin{minipage}{0.98\linewidth}
\footnotesize
\textit{Notes.} $\Delta \mathrm{CFI}_{t}(k)=\mathrm{CFI}_{t}^{\text{full}}-\mathrm{CFI}_{t}^{(-k)}$. Targeted removes the top-$k$ categories by date-specific RCS. Strength removes the top-$k$ categories by node strength. Random reports the Monte Carlo mean across dates.
\end{minipage}
\end{table}

\begin{figure}[t]
    \centering
    \includegraphics[width=0.98\linewidth]{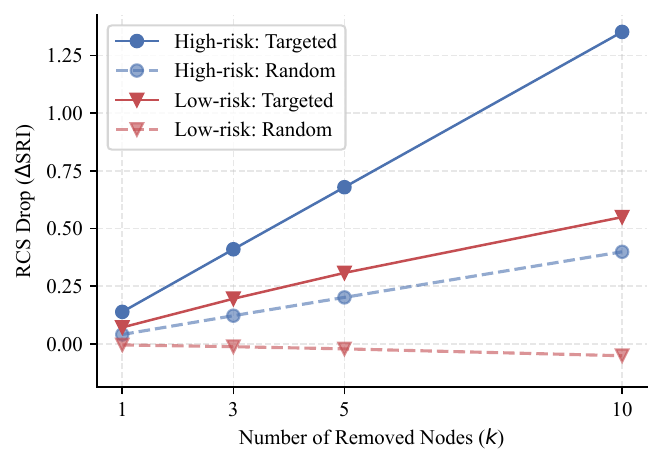}
    \caption{Attack curves in high-fragility vs.\ low-fragility regimes (defined by top/bottom CFI quantiles). Targeted removal is markedly more effective in high-fragility states.}
    \label{fig:attack_hilo_rcs}
\end{figure}

Finally, we repeat the attack tests by across rolling windows, conditioning on system state by splitting the sample into high- and low-fragility regimes defined by the top and bottom 20\% of the CFI distribution (48 windows each). Fig.~\ref{fig:attack_hilo_rcs} shows that targeted removal is markedly more effective in high-fragility windows consistent with the interpretation that stressed states are more sensitive to targeted de-risking.

Overall, the RCS provides a structural decomposition of ecosystem-wide fragility by identifying protocol types whose position in the dependence network sustains high CFI states. Empirically, structurally important categories are not necessarily the largest by TVL, and targeted removal based on RCS yields substantially larger counterfactual fragility reductions than random removal, especially during high-fragility regimes. These findings support the use of RCS as a node-level tool for structural stress testing and scenario analysis that complements the system-level perspective offered by the CFI.

\section{Robustness Analysis}
\label{sec:robustness}
\noindent
This section examines the robustness of the proposed network-based fragility measures to alternative modeling choices. We focus on two dimensions that are most likely to affect inference in correlation-network settings: (i) dependence estimation methods, and (ii) metric construction and threshold choices.

\subsection{Alternative Network Specifications}
\label{sec:robustness_alt_spec}
\noindent
We first assess robustness to alternative dependence estimators. In addition to the baseline Ledoit--Wolf shrinkage correlation, we re-estimate rolling networks using sample Pearson correlations and partial correlations estimated via Graphical LASSO.

\begin{figure}[t]
    \centering
    \includegraphics[width=\linewidth]{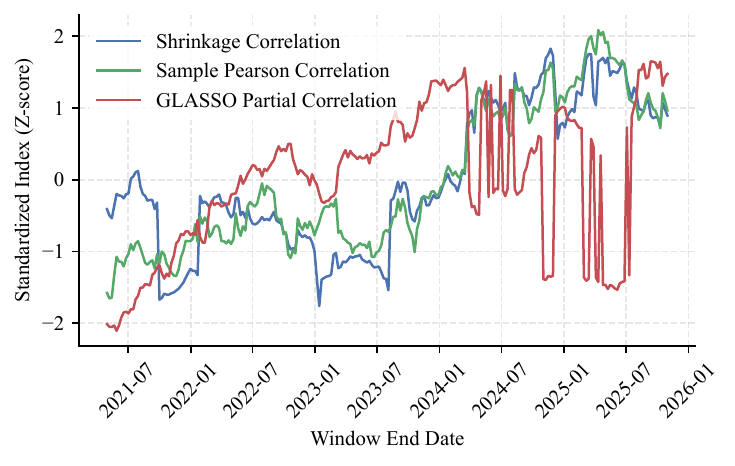}
    \caption{Alternative network specifications: standardized co-movement proxy. The figure compares z-scored co-movement proxies based on shrinkage correlation, sample Pearson correlation, and Graphical LASSO partial-correlations. Shrinkage- and sample-based measures track each other closely, while the Glasso-based proxy differs due to its focus on conditional dependence.}
    \label{fig:alt_spec_comovement_main}
\end{figure}

Fig.~\ref{fig:alt_spec_comovement_main} compares the standardized time series of the primary co-movement proxy across the three specifications. The shrinkage- and sample-correlation measures exhibit highly synchronized dynamics throughout the sample, confirming that aggregate synchronization is not driven by the specific correlation estimator. By contrast, the Glasso-based (partial-correlation) proxy deviates substantially from correlation-based measures. This divergence is expected: partial correlations remove common market-wide components and therefore capture conditional linkages rather than overall synchronization. Consequently, partial-correlation networks provide complementary structural information but are not substitutes for correlation-based systemic fragility indicators.

\subsection{Threshold Sensitivity}
\label{sec:robust_threshold}
\begin{figure}[t]
    \centering
    \includegraphics[width=\linewidth]{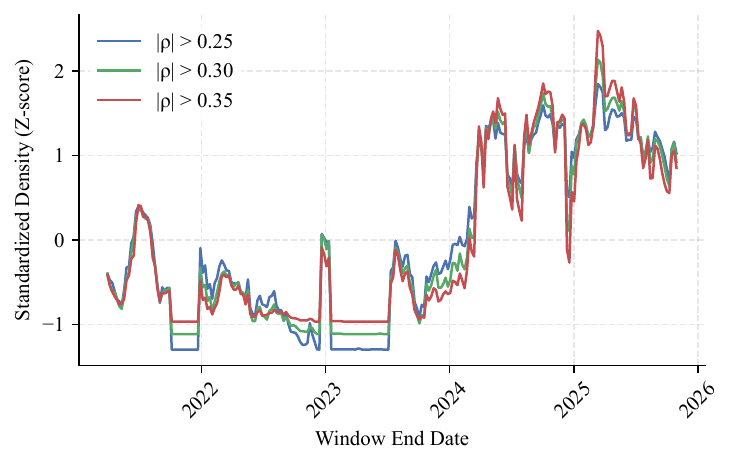}
    \caption{Threshold sensitivity of strong-edge density under alternative correlation cutoffs
    ($|\rho|>0.25, 0.30, 0.35$). The strong-edge density dynamics remain stable across thresholds.}
    \label{fig:threshold_sensitivity_main}
\end{figure}

\noindent
Finally, we assess sensitivity to the correlation cutoff used to define strong edges in the strong-edge density metric. We recompute strong-edge density using $|\rho|>0.25$, $0.30$ (baseline), and $0.35$. Fig.~\ref{fig:threshold_sensitivity_main} shows that the resulting standardized series are highly stable across these reasonable choices, implying that the timing and relative intensity of identified high-fragility periods are not driven by an arbitrary threshold.

Taken together, these robustness checks confirm that the main conclusions are not driven by a particular estimator, node universe definition, or ad hoc parameter setting. Instead, they reflect persistent and economically meaningful changes in the evolving correlation structure of the DeFi ecosystem.

\section{Discussion}
\label{sec:discussion}
\noindent
This paper studies systemic risk in decentralized finance from a network perspective, focusing on the evolving dependence structure across protocol types. The proposed CFI captures a dimension of systemic risk that is structural rather than price-based. Empirically, the CFI is weakly related to asset-level volatility, yet closely aligned with instability in aggregate liquidity. This distinction highlights that network fragility reflects synchronization and loss of diversification across functional modules, rather than short-term market fluctuations.

The dynamics of the CFI suggest that systemic vulnerability in DeFi accumulates gradually through endogenous synchronization. High-fragility states are persistent and coincide with increasing concentration in network dependence, supporting the view that systemic risk emerges as a slow-moving structural condition rather than only as a response to discrete shocks. This highlights the value of network-based indicators for continuous risk monitoring, complementing event-driven analyses commonly used in the literature.

At the node level, the RCS shows that systemic importance is not determined by economic size. Protocol categories that sustain high-fragility states are often structurally central in the dependence network despite modest TVL. Moreover, heightened systemic risk is associated with greater participation in synchronized behavior than concentration in a small core, underscoring the ecosystem-wide nature of fragility in DeFi.

Finally, counterfactual attack tests demonstrate that RCS-based targeted removal yields significantly larger reductions in system-level fragility than random interventions, particularly during high-fragility regimes. In conclusion, these results emphasize that systemic risk in DeFi is best understood as a structural network phenomenon, for which the CFI and RCS provide complementary tools for monitoring and stress testing.

\section{Conclusion}
\label{sec:conclusion}
\noindent
This paper develops a network-based framework to measure systemic risk in decentralized finance using time-varying correlation networks from category-level TVL dynamics. We introduce the CFI as a measure of ecosystem-wide structural vulnerability, and the RCS to decompose this vulnerability across protocol types based on marginal structural impact.

Empirically, the framework reveals that systemic fragility in DeFi is driven by the gradual synchronization of dependence structures rather than short-term price volatility. Periods of elevated CFI coincide with increased instability in aggregate liquidity, even when market volatility remains subdued. At the node level, systemically important protocol categories are identified not by economic size but by their persistent structural positioning within the correlation network, highlighting a distinction between large and systemically important actors.

By linking network-level fragility with node-level contributions, the proposed framework provides a coherent and interpretable tool for monitoring systemic risk in DeFi. More broadly, the results underscore the importance of structural dependence as a channel of systemic risk in decentralized markets, with implications for risk management, protocol design, and macroprudential oversight.

Several avenues for future research and application emerge naturally. One extension is to compare the CFI with established connectedness measures from the financial econometrics literature, to contrast correlation-based network fragility with alternative, model-based notions of systemic connectedness. Extending the framework to high-frequency data and incorporating cross-chain bridge activity would also help characterize the propagation of systemic risk across rapidly evolving blockchain ecosystems. From a practical perspective, the CFI and RCS provide a blueprint for DeFi-native macroprudential tools that adjust risk parameters in response to elevated systemic connectivity.

\ifnum\ANON=1
    \section*{Acknowledgment}
    \noindent The authors acknowledge the use of generative AI for manuscript development. Specifically, ChatGPT-5 (OpenAI) was used for editing and grammar enhancement. All content generated by the AI was reviewed, edited, and verified for accuracy by the human authors.

\else
    \section*{Acknowledgment}
    \noindent SZ's PhD is funded by a China Scholarship Council/University of Bristol joint scholarship, No. 202410320012. 
    JC is partly funded by UK Research and Innovation (UKRI) Engineering and Physical Sciences Research Council (EPSRC) [grant number EP/Y028392/1]: AI for Collective Intelligence (AI4CI). The authors acknowledge the use of generative AI for manuscript development. Specifically, ChatGPT-5 (OpenAI) was used for editing and grammar enhancement. All content generated by the AI was reviewed, edited, and verified for accuracy by the human authors.
\fi

\balance
\bibliographystyle{IEEEtran}
\bibliography{reference}

\begin{thebibliography}{10}
\providecommand{\url}[1]{#1}
\csname url@samestyle\endcsname
\providecommand{\newblock}{\relax}
\providecommand{\bibinfo}[2]{#2}
\providecommand{\BIBentrySTDinterwordspacing}{\spaceskip=0pt\relax}
\providecommand{\BIBentryALTinterwordstretchfactor}{4}
\providecommand{\BIBentryALTinterwordspacing}{\spaceskip=\fontdimen2\font plus
\BIBentryALTinterwordstretchfactor\fontdimen3\font minus \fontdimen4\font\relax}
\providecommand{\BIBforeignlanguage}[2]{{%
\expandafter\ifx\csname l@#1\endcsname\relax
\typeout{** WARNING: IEEEtran.bst: No hyphenation pattern has been}%
\typeout{** loaded for the language `#1'. Using the pattern for}%
\typeout{** the default language instead.}%
\else
\language=\csname l@#1\endcsname
\fi
#2}}
\providecommand{\BIBdecl}{\relax}
\BIBdecl

\bibitem{leon2025data}
\BIBentryALTinterwordspacing
J.~C. Le{\'o}n and A.~Lehar, ``What data have told us about decentralized finance,'' \emph{Journal of Corporate Finance}, p. 102916, 2025. [Online]. Available: \url{https://doi.org/10.1016/j.jcorpfin.2025.102916}
\BIBentrySTDinterwordspacing

\bibitem{schar2021decentralized}
\BIBentryALTinterwordspacing
F.~Sch{\"a}r, ``Decentralized finance: On blockchain-and smart contract-based financial markets,'' \emph{Federal Reserve Bank of St. Louis Review}, vol. 103, no.~2, pp. 153--174, 2021. [Online]. Available: \url{https://dx.doi.org/10.20955/r.103.153-74}
\BIBentrySTDinterwordspacing

\bibitem{werner2022sok}
\BIBentryALTinterwordspacing
S.~Werner, D.~Perez, L.~Gudgeon, A.~Klages-Mundt, D.~Harz, and W.~Knottenbelt, ``Sok: Decentralized finance ({DeFi}),'' in \emph{Proceedings of the 4th ACM Conference on Advances in Financial Technologies}, 2022, pp. 30--46. [Online]. Available: \url{https://dl.acm.org/doi/abs/10.1145/3558535.3559780}
\BIBentrySTDinterwordspacing

\bibitem{board2023financial}
\BIBentryALTinterwordspacing
{Financial Stability Board}, ``The financial stability risks of decentralised finance,'' Financial Stability Board: Basel, Switzerland, Tech. Rep., 16 Feb 2023. [Online]. Available: \url{https://www.fsb.org/uploads/P160223.pdf}
\BIBentrySTDinterwordspacing

\bibitem{haldane2011systemic}
\BIBentryALTinterwordspacing
A.~G. Haldane and R.~M. May, ``Systemic risk in banking ecosystems,'' \emph{Nature}, vol. 469, no. 7330, pp. 351--355, 2011. [Online]. Available: \url{https://doi.org/10.1038/nature09659}
\BIBentrySTDinterwordspacing

\bibitem{bardoscia2017pathways}
\BIBentryALTinterwordspacing
M.~Bardoscia, S.~Battiston, F.~Caccioli, and G.~Caldarelli, ``Pathways towards instability in financial networks,'' \emph{Nature Communications}, vol.~8, no.~1, p. 14416, 2017. [Online]. Available: \url{https://doi.org/10.1038/ncomms14416}
\BIBentrySTDinterwordspacing

\bibitem{mantegna1999hierarchical}
\BIBentryALTinterwordspacing
R.~N. Mantegna, ``Hierarchical structure in financial markets,'' \emph{The European Physical Journal B-Condensed Matter and Complex Systems}, vol.~11, no.~1, pp. 193--197, 1999. [Online]. Available: \url{https://doi.org/10.1007/s100510050929}
\BIBentrySTDinterwordspacing

\bibitem{billio2012econometric}
\BIBentryALTinterwordspacing
M.~Billio, M.~Getmansky, A.~W. Lo, and L.~Pelizzon, ``Econometric measures of connectedness and systemic risk in the finance and insurance sectors,'' \emph{Journal of Financial Economics}, vol. 104, no.~3, pp. 535--559, 2012. [Online]. Available: \url{https://doi.org/10.1016/j.jfineco.2011.12.010}
\BIBentrySTDinterwordspacing

\bibitem{diebold2014network}
\BIBentryALTinterwordspacing
F.~X. Diebold and K.~Y{\i}lmaz, ``On the network topology of variance decompositions: Measuring the connectedness of financial firms,'' \emph{Journal of Econometrics}, vol. 182, no.~1, pp. 119--134, 2014. [Online]. Available: \url{https://doi.org/10.1016/j.jeconom.2014.04.012}
\BIBentrySTDinterwordspacing

\bibitem{acemoglu2015systemic}
\BIBentryALTinterwordspacing
D.~Acemoglu, A.~Ozdaglar, and A.~Tahbaz-Salehi, ``Systemic risk and stability in financial networks,'' \emph{American Economic Review}, vol. 105, no.~2, pp. 564--608, 2015. [Online]. Available: \url{https://doi.org/10.1257/aer.20130456}
\BIBentrySTDinterwordspacing

\bibitem{battiston2012debtrank}
\BIBentryALTinterwordspacing
S.~Battiston, M.~Puliga, R.~Kaushik, P.~Tasca, and G.~Caldarelli, ``{DebtRank}: Too central to fail? {F}inancial networks, the {FED} and systemic risk,'' \emph{Scientific Reports}, vol.~2, no.~1, p. 541, 2012. [Online]. Available: \url{https://doi.org/10.1038/srep00541}
\BIBentrySTDinterwordspacing

\bibitem{aquilina2025cryptocurrencies}
\BIBentryALTinterwordspacing
M.~Aquilina, G.~Cornelli, J.~Frost, and L.~Gambacorta, ``Cryptocurrencies and decentralised finance: functions and financial stability implications,'' Bank for International Settlements, BIS Papers 156, Apr 2025. [Online]. Available: \url{https://www.bis.org/publ/bppdf/bispap156.pdf}
\BIBentrySTDinterwordspacing

\bibitem{kitzler2023disentangling}
\BIBentryALTinterwordspacing
S.~Kitzler, F.~Victor, P.~Saggese, and B.~Haslhofer, ``Disentangling decentralized finance {(DeFi)} compositions,'' \emph{ACM Transactions on the Web}, vol.~17, no.~2, pp. 1--26, 2023. [Online]. Available: \url{https://doi.org/10.1145/3532857}
\BIBentrySTDinterwordspacing

\bibitem{qin2021attacking}
\BIBentryALTinterwordspacing
K.~Qin, L.~Zhou, B.~Livshits, and A.~Gervais, ``Attacking the {DeFi} ecosystem with flash loans for fun and profit,'' in \emph{Financial Cryptography and Data Security. FC21. Lecture Notes in Computer Science}.\hskip 1em plus 0.5em minus 0.4em\relax Berlin: Springer, 2021, vol. 12674, pp. 3--32. [Online]. Available: \url{https://doi.org/10.1007/978-3-662-64322-8_1}
\BIBentrySTDinterwordspacing

\bibitem{auer2018regulating}
R.~Auer and S.~Claessens, ``Regulating cryptocurrencies: assessing market reactions,'' Bank for International Settlements, BIS Quarterly Review, Sep 2018.

\bibitem{fakhfekh2024dependence}
\BIBentryALTinterwordspacing
M.~Fakhfekh, A.~Bejaoui, A.~F. Bariviera, and A.~Jeribi, ``Dependence structure between {NFT}, {DeFi} and cryptocurrencies in turbulent times: An archimax copula approach,'' \emph{The North American Journal of Economics and Finance}, vol.~70, p. 102079, 2024. [Online]. Available: \url{https://doi.org/10.1016/j.najef.2024.102079}
\BIBentrySTDinterwordspacing

\bibitem{feng2025research}
\BIBentryALTinterwordspacing
X.~Feng, M.~Yu, T.~Yan, J.~Lin, and C.~J. Tessone, ``Research on the time-varying network topology characteristics of cryptocurrencies on {U}niswap v3,'' \emph{Electronics}, vol.~14, no.~12, p. 2444, 2025. [Online]. Available: \url{https://doi.org/10.3390/electronics14122444}
\BIBentrySTDinterwordspacing

\bibitem{yan2025network}
\BIBentryALTinterwordspacing
T.~Yan and C.~J. Tessone, ``Network analysis of {U}niswap: Centralization and fragility in the decentralized exchange market,'' in \emph{Proceedings of Blockchain Kaigi 2024 (BCK24)}.\hskip 1em plus 0.5em minus 0.4em\relax Physical Society of Japan, 2025, p. 011013. [Online]. Available: \url{https://journals.jps.jp/doi/abs/10.7566/JPSCP.44.011013}
\BIBentrySTDinterwordspacing

\bibitem{wu2025dexposure}
\BIBentryALTinterwordspacing
W.~Wu, K.~Qian, A.~Lui, C.~Jack, Y.~Wu, P.~McBurney, F.~He, and B.~Zhang, ``{DeXposure}: A dataset and benchmarks for inter-protocol credit exposure in decentralized financial networks,'' 2025. [Online]. Available: \url{https://arxiv.org/abs/2511.22314}
\BIBentrySTDinterwordspacing

\bibitem{tovanich2023contagion}
\BIBentryALTinterwordspacing
N.~Tovanich, M.~Kassoul, S.~Weidenholzer, and J.~Prat, ``Contagion in decentralized lending protocols: A case study of compound,'' in \emph{Proceedings of the 2023 Workshop on Decentralized Finance and Security}, 2023, pp. 55--63. [Online]. Available: \url{https://dl.acm.org/doi/abs/10.1145/3605768.3623544}
\BIBentrySTDinterwordspacing

\bibitem{ledoit2004well}
\BIBentryALTinterwordspacing
O.~Ledoit and M.~Wolf, ``A well-conditioned estimator for large-dimensional covariance matrices,'' \emph{Journal of Multivariate Analysis}, vol.~88, no.~2, pp. 365--411, 2004. [Online]. Available: \url{https://doi.org/10.1016/S0047-259X(03)00096-4}
\BIBentrySTDinterwordspacing

\bibitem{demiguel2009optimal}
\BIBentryALTinterwordspacing
V.~DeMiguel, L.~Garlappi, and R.~Uppal, ``Optimal versus naive diversification: How inefficient is the 1/n portfolio strategy?'' \emph{The Review of Financial Studies}, vol.~22, no.~5, pp. 1915--1953, 2009. [Online]. Available: \url{https://doi.org/10.1093/rfs/hhm075}
\BIBentrySTDinterwordspacing

\bibitem{ledoit2017nonlinear}
\BIBentryALTinterwordspacing
O.~Ledoit and M.~Wolf, ``Nonlinear shrinkage of the covariance matrix for portfolio selection: {M}arkowitz meets {G}oldilocks,'' \emph{The Review of Financial Studies}, vol.~30, no.~12, pp. 4349--4388, 2017. [Online]. Available: \url{https://doi.org/10.1093/rfs/hhx052}
\BIBentrySTDinterwordspacing

\bibitem{engle2002dynamic}
\BIBentryALTinterwordspacing
R.~Engle, ``Dynamic conditional correlation: A simple class of multivariate generalized autoregressive conditional heteroskedasticity models,'' \emph{Journal of Business \& Economic Statistics}, vol.~20, no.~3, pp. 339--350, 2002. [Online]. Available: \url{https://doi.org/10.1198/073500102288618487}
\BIBentrySTDinterwordspacing

\bibitem{forbes2002no}
\BIBentryALTinterwordspacing
K.~J. Forbes and R.~Rigobon, ``No contagion, only interdependence: measuring stock market comovements,'' \emph{The Journal of Finance}, vol.~57, no.~5, pp. 2223--2261, 2002. [Online]. Available: \url{https://doi.org/10.1111/0022-1082.00494}
\BIBentrySTDinterwordspacing

\bibitem{onnela2003dynamics}
\BIBentryALTinterwordspacing
J.-P. Onnela, A.~Chakraborti, K.~Kaski, J.~Kertesz, and A.~Kanto, ``Dynamics of market correlations: Taxonomy and portfolio analysis,'' \emph{Physical Review E}, vol.~68, no.~5, p. 056110, 2003. [Online]. Available: \url{https://doi.org/10.1103/PhysRevE.68.056110}
\BIBentrySTDinterwordspacing

\bibitem{tumminello2007correlation}
\BIBentryALTinterwordspacing
M.~Tumminello, T.~Di~Matteo, T.~Aste, and R.~N. Mantegna, ``Correlation based networks of equity returns sampled at different time horizons,'' \emph{The European Physical Journal B}, vol.~55, no.~2, pp. 209--217, 2007. [Online]. Available: \url{https://doi.org/10.1140/epjb/e2006-00414-4}
\BIBentrySTDinterwordspacing

\bibitem{barrat2004architecture}
\BIBentryALTinterwordspacing
A.~Barrat, M.~Barthelemy, R.~Pastor-Satorras, and A.~Vespignani, ``The architecture of complex weighted networks,'' \emph{Proceedings of the National Academy of Sciences}, vol. 101, no.~11, pp. 3747--3752, 2004. [Online]. Available: \url{https://doi.org/10.1073/pnas.0400087101}
\BIBentrySTDinterwordspacing

\bibitem{laloux1999noise}
\BIBentryALTinterwordspacing
L.~Laloux, P.~Cizeau, J.-P. Bouchaud, and M.~Potters, ``Noise dressing of financial correlation matrices,'' \emph{Physical Review Letters}, vol.~83, no.~7, p. 1467, 1999. [Online]. Available: \url{https://doi.org/10.1103/PhysRevLett.83.1467}
\BIBentrySTDinterwordspacing

\bibitem{plerou2002random}
\BIBentryALTinterwordspacing
V.~Plerou, P.~Gopikrishnan, B.~Rosenow, L.~A.~N. Amaral, T.~Guhr, and H.~E. Stanley, ``Random matrix approach to cross correlations in financial data,'' \emph{Physical Review E}, vol.~65, no.~6, p. 066126, 2002. [Online]. Available: \url{https://doi.org/10.1103/PhysRevE.65.066126}
\BIBentrySTDinterwordspacing

\bibitem{tumminello2005tool}
\BIBentryALTinterwordspacing
M.~Tumminello, T.~Aste, T.~Di~Matteo, and R.~N. Mantegna, ``A tool for filtering information in complex systems,'' \emph{Proceedings of the National Academy of Sciences}, vol. 102, no.~30, pp. 10\,421--10\,426, 2005. [Online]. Available: \url{https://doi.org/10.1073/pnas.0500298102}
\BIBentrySTDinterwordspacing

\bibitem{kwapien2012physical}
\BIBentryALTinterwordspacing
J.~Kwapie{\'n} and S.~Dro{\.z}d{\.z}, ``Physical approach to complex systems,'' \emph{Physics Reports}, vol. 515, no. 3-4, pp. 115--226, 2012. [Online]. Available: \url{https://doi.org/10.1016/j.physrep.2012.01.007}
\BIBentrySTDinterwordspacing

\bibitem{hollo2012ciss}
\BIBentryALTinterwordspacing
D.~Hollo, M.~Kremer, and M.~Lo~Duca, ``{CISS} -- {A} composite indicator of systemic stress in the financial system,'' European Central Bank, Working Paper Series 1426, 2012. [Online]. Available: \url{https://www.ecb.europa.eu/pub/pdf/scpwps/ecbwp1426.pdf}
\BIBentrySTDinterwordspacing

\end{thebibliography}

\end{document}